\documentclass[12pt]{article}
\title{ Unusual condensates in quark and atomic systems\thanks{
Lectures delivered at 8th Moscow school of Physics (33rd ITEP
Winter School of Physics)}
\\}
\author{B.Kerbikov,\thanks{E-mail:borisk@itep.ru}\\
 State Research
Center\\Institute of Theoretical and Experimental Physics, \\
Moscow, 117218, Russia }
 \date{}
  \newcommand{\beeq}{\begin{eqnarray}}
\newcommand{\eeq}{\end{eqnarray}}
 \newcommand{\be}{\begin{equation}}
\newcommand{\ee}{\end{equation}}
\def\la{\mathrel{\mathpalette\fun
<}} \def\ga{\mathrel{\mathpalette\fun >}}
\def\fun#1#2{\lower3.6pt\vbox{\baselineskip0pt\lineskip.9pt
\ialign{$\mathsurround=0pt#1\hfil ##\hfil$\crcr#2\crcr\sim\crcr}}}

\newcommand{\vek}{\mbox{\boldmath${\rm k}$}}

\newcommand{\ver}{\mbox{\boldmath${\rm r}$}}

\newcommand{\vep}{\mbox{\boldmath${\rm p}$}}

\newcommand{\lan}{\langle}
\newcommand{\ran}{\rangle}

\begin{document}
\maketitle

\begin{abstract}

In these lectures we discuss condensates which are formed in quark
matter when it is squeezed and in a gas of fermionic atoms when it
is cooled. The behavior of these two seemingly very different
systems reveals striking similarities. In particular, in both
systems the Bose-Einstein condensate to Bardeen--Cooper-Schrieffer
(BEC-BCS) crossover takes place.

\end{abstract}
\section{Introduction}

~~~~~In these lectures I will review the striking discoveries of
the new phases of the quark and atomic matter. Squeezing the quark
matter and cooling the gas of fermionic atoms lead to the
formation of condensates. Condensates in both systems exhibit
similar properties. The most interesting point is the crossover
between  different regimes of condensation, which takes place
under the variation of certain  external parameters.

At first sight the dense quark matter and cold fermionic atoms
have nothing in common as one might conclude from Table~1.

\begin{center}
{\bf Table 1}\\

Comparison of density and temperature in  quark  and atomic
systems\footnote{We remind that 1 eV =11606 K}

\vspace{0.5cm}

%
\begin{tabular}{|l|l|l|}
\hline
~&&\\
  &  density& temperature \\
  ~&&\\
\hline
~&&\\
 quarks& $> 10^{39} cm^{-3}$& $\la 5\cdot 10^{11} K$\\
 ~&&\\
 \hline
 ~&&\\
 fermionic  atoms& $(10^{13} -10^{14}) cm^{-3}$& $\sim (10^{-7}-10^{-8})K$\\
 ~&&\\
 \hline

\end{tabular}
\end{center}


The numbers presented in Table 1 will be discussed below, but some
comments are in order already at this point. The quark density
$n\sim 10^{39} cm ^{-3}$ is not high, it is only three times
larger than the density of quarks in normal nuclear matter. When
the density reaches this value the gap equation develops a
nontrivial solution , which was commonly interpreted  as the onset
of the BCS-like phase. We shall argue that this is a misleading
conclusion and the BCS regime corresponds to much higher
densities.

The density of fermionic atoms $n\sim (10^{13}-10^{14}) cm^{-3}$
is very low, it is about five orders of magnitude lower than the
density of the air. Such a low density is needed in order  to
suppress the formation of molecules and clusters. To obtain
quantum condensation in such a dilute system it should  be cooled
to extremely low temperatures $\sim (10^{-7} - 10^{-8})K$ since
roughly speaking the temperature of BEC scales as $n^{2/3}$. The
fascinating aspect about ultracold fermionic atoms is that the
strength of the interatomic  interaction can be tuned using the
magnetic Feshbach resonance (see below).

Intense discussion of the unusual properties of quark matter at
finite density (color superconductivity) has started in 1998 after
the publication of the pioneering papers \cite{1,2}. Condensates
in a gas of ultracold fermionic atoms have been experimentally
created in 2003 \cite{3,4,5}. The idea that the BCS regime in
quark system sets in as a result of the BEC-BCS crossover was
formulated in 2001 \cite{6,7}. Hundreds of papers have been
published in recent years on quark systems at nonzero density and
on ultracold fermionic atoms. The reader willing to get deeper and
broader knowledge on both subjects may address review papers
\cite{8,9}.

\section{Two types of condensation: BEC and BCS}

~~~~~It is not easy to give a universal definition of a
condensate, through in any concrete situation this notion is well
understood. The key properties of a condensate are:
\begin{description}
  \item[(a)] Long-range order.
  \item[(b)] Macroscopic quantum coherence, i.e., the possibility
  to regard the condensate as a single quantum entity described
  by the macroscopic wave function.
\end{description}

The type of condensate (the character of quantum degeneracy)
depends upon the spin of the particles. In a system of bosons BEC
takes place when de Broglie thermal wavelength $\lambda_B = \left(
2\pi/mT\right)^{1/2}$ is of the order of interparticle distance,
$\lambda_B n^{1/3} \sim 1$. More accurate  analysis yields the
degeneracy condition in the form $\lambda^3_Bn=s\zeta(3/2)$, or
\be T_c({\rm BEC}) =\frac{2\pi}{m} \left [
\frac{n}{s\zeta\left(\frac32\right)}\right]^{2/3}.\label{u1}\ee

Here  $s$ is the number of possible spin states, $ \zeta(3/2)
\simeq 2.6$ is the Riemann zeta function, and  we set $k_B=\hbar
=1$.

BEC has a very long history. The best known example of this regime
is $~^4He$ below the so-called $\lambda$-point at 2.17 K.
Condensations of $\pi$- and $K$- mesons ware discussed as possible
states of neutron  stars. BEC-like behavior is present in some
scenarios of high $T_c$ superconductivity. Ten years ago BEC was
observed in  trapped gases of bosonic alkali atoms $~^{87}Rb$ and
$~^{23}Na$ \cite{10}-\cite{12}. In these lectures we shall discuss
BEC and BCS condensation in the systems  of fermionic atoms  and
quarks. A natural question at this point would be : "How is it
possible to talk about BEC for fermions?" The answer to this
question is the following. Interaction between fermions may  lead
to pairing  and the resulting composite  bosons may  undergo BEC
-- "A molecular Bose--Einstein condensate emerges from a Fermi
sea" \cite{3}. This was realized experimentally for two fermionic
alkalies -- $~^6Li$ and $~^{40}K$ \cite{3,4,5}. Both have even
number of nucleons and one electron outside  the  closed shell and
therefore they are fermions. For  $~^6Li_2$ molecules with the
density $n=10^{13}$ cm$^{-3}$ Eq.(1) gives $T_c\simeq 5\cdot
10^{-7}K$ in a fair agreement with the experimental result
$T_c=2.8\cdot 10^{-7} K$ \cite{4}. In reality, however, (1) is not
applicable to the trapped gases. Condensation in this case means
macroscopic population of the lowest trap  state and the critical
temperature is given by $T_c=\omega \left(N/\zeta(3)\right)^{1/3}
\simeq 0.94 \omega N^{1/3} $ \cite{13}, where $N$ is the number of
particles in the trap (typically $N\sim 10^5)$,  $\omega$ is the
average oscillation frequency of the trap  ($\omega\sim $ few
$nK$), $\zeta(3) \simeq1.2$. Thermodynamic limit  corresponds to
$N\to \infty, \omega \to \infty, N\omega^3 =const$ \cite{13}.
Other important factors making the thermodynamic of the trapped
gas nontrivial are the trap geometry, inhomogeneity of the system,
and the effects of the interaction between the molecules
\cite{13}-\cite{16}.

The observation  of BEC for fermionic atoms is not the main reason
why the experiments \cite{3,4,5} generated enormous excitement.
The striking discovery was the creation of a state with another
type of pairing between the atoms, namely the BCS condensate and
the observation  of the crossover between BEC and BCS regimes
[3-5,9]. Before turning to the crossover phenomena let us remind
few basic facts about Cooper pairing and BCS.

Cooper pairing is the cornerstone of the theory of
superconductivity -- one of the most elegant theories created in
the last century. The nature of Cooper pairing  is different from
that of molecular-like states discussed  above\footnote{Compact
electron pairs were introduced \cite{17} in an attempt to explain
superconductivity couple of years before the BCS theory was
formulated. Sometimes compact molecular-like states are called
Schafroth pairs.}. The size of the Cooper pair is characterized by
the coherence length $\xi$ which is macroscopic, i.e. much larger
than the interatomic distance, $\xi\sim 10^{-4}$ cm, while $a\sim
10^{-8}$ cm. The important dimensionless parameter is
$n^{1/3}\xi$, where $n$ is the density of fermions. For a typical
superconductor the density of electrons is $n\sim 10^{22}$
cm$^{-3}$, so that $n^{1/3}\xi \ga 10^3$. Alternatively, instead
of $n^{1/3}$ we may use the Fermi momentum $k_F=(3\pi^3n)^{1/3}$.
Thus the BCS regime is characterized by the dimensionless
parameter \be k_F\xi \ga 10^{3}.\label{2}\ee Equation (\ref{2})
means that Cooper pairs strongly overlap with each other and the
problem is essentially many-body. In coordinate space the wave
function of the Cooper pair is  proportional to $(\sin
k_Fr/k_Fr)\exp(-r/\xi)$ and therefore it has $\ga 10^3$ nodes. We
see that pairing in the BCS regime is very different from the
usual molecular-like one.  The dynamical mechanism leading to the
Cooper phenomenon includes three important elements:
\begin{description}
 \item[(i)] Attraction between particles.

  \item[(ii)]  Existence of the the Fermi surface.
  \item[(iii)] Interaction must be concentrated within a thin layer of
momentum space around the Fermi surface.

\end{description}

Attraction between electrons in a  superconductor is due to the
interaction with the lattice (phonon mechanism). The
characteristic energy scale of this interaction is the Debye
frequency $\omega_D \simeq \frac12 \left(m/M\right)^{1/2} Ry \sim
10^{-2}$ eV, where  $m$ and $M$ are the electron and ion,
masses~~\footnote{Close to the Fermi surface this attraction
overcomes Coulomb repulsion -- see, e.g. \, \cite{18}.}. On the
other hand, the typical Fermi energy is $E_F=k^2_F/2m \simeq
(3\pi^2 n)^{2/3}/2m \simeq 2$ eV. Thus $\omega_D\ll E_F$ and we
see that the interaction in the BCS regime  is really concentrated
within a thin layer around the Fermi surface. Electrons interact
in two dimensions instead of three, the sum $\sum_k \vek^{-2}$
logarithmically diverges at small momenta leading the Cooper
instability and to the  gap in the spectrum.

At this point we wish to demonstrate the importance of  the point
(iii), or the hierarchy $\omega_D\ll E_F$, in a more formal way.
Use will be made of the technique standard for  the  quantum field
theory. Similar language was used, e.g., in Refs. \cite{19,20}. We
start with the partition function in the form of the path integral
over the fermion fields $$Z=tr ~e^{-\beta H} =\int D\psi^+
D\psi\times $$ \be \times \exp \left\{ -\int^\beta_0 d\tau \int
d\ver \left[ \sum_\sigma \psi^+_\sigma
\left(\frac{\partial}{\partial
\tau}-\frac{\nabla^2}{2m}-\mu\right) \psi_\sigma- g \psi^+_\sigma
\psi^+_{-\sigma}
\psi_{-\sigma}\psi_\sigma\right]\right\}.\label{3}\ee

Here $\sigma$ is the spin variable, $\mu\simeq E_F$ due to
(iii)\footnote{We stress that $\mu\simeq E_F$ is true only in the
BCS regime.}. The constant $g$ has the dimension of $m^{-2}$, the
dimension of $\psi_\sigma(\ver, \tau)$ is $m^{3/2}$. From now on
we shall consider stationary homogeneous superconductor and assume
that the four-fermion interaction is local. The last term in the
exponent in (\ref{3}) is quartic in the fermion fields making the
direct integration impossible. This difficulty can be overcome by
Hubbard--Stratonovich trick (bosonization). To decouple the
quartic interaction one writes $$ \exp \left\{ g\int^\beta_0 d\tau
\int d\ver \psi^+_\sigma\psi^+_{-\sigma} \psi_{-\sigma}
\psi_\sigma\right\}=$$ \be = \int Dd^*Dd\exp \left\{ -\int^\beta_0
d\tau \int d\ver \left(\frac{|d|^2}{g} -d\psi^+_{\sigma}
\psi^+_{-\sigma}- d^*\psi_{-\sigma}
\psi_\sigma\right)\right\}.\label{4}\ee

Here we have  introduced complex  scalar field $d$. The quantity
$|d|$ will acquire the meaning of the superconducting order
parameter. From the Lagrange equation of motion for $d^*$ we see
that $d=g\psi_{-\sigma} \psi_\sigma$, so that $d$ has a  dimension
of mass as it should be for the  gap. Before we insert (\ref{4})
into (\ref{3})  and proceed  on with our derivation of $Z$, some
important remarks should be made.

The initial microscopic Hamiltonian in (\ref{3}) is proportional
to $\psi^+_\sigma\psi^+_{-\sigma} \psi_{-\sigma} \psi_\sigma$.
According to Wick's  theorem it has to be factorized as $\lan
\psi^+_\sigma\psi_{-\sigma} \ran \lan \psi^+_{-\sigma}
\psi_\sigma\ran$. In such a form the Hamiltonian conserves the
number of particles and does not  lead to superconductivity.
Equation (\ref{4}) corresponds to an alternative factorization
$\lan \psi^+_\sigma\psi^+_{-\sigma}\ran \lan \psi_{-\sigma}
\psi_\sigma\ran$. In a normal, non-condensed state such anomalous
averages  correspond to the off-diagonal long range order (ODLRO)
-- a concept introduced by Yang \cite{21}, and relevant both for
BEC and BCS regimes. The state with ODLRO is characterized by a
long range correlation

\be S(\ver_1, \ver_2) \propto d^* (\ver_1) d (\ver_2),
~~|\ver_1-\ver_2|\to \infty.\label{5}\ee

The wave function in ODLRO regime is a coherent state with
fluctuating number of particles \cite{22,23} \be |\Psi\ran =\exp
\left\{ -\frac{N}{2} +\sqrt{N} b^+\right\} |0\ran,\label{6}\ee
where $b^+$ is a pair creation operator. For a metallic
superconductor $N\sim 10^{20}$ and  the uncertainty in the number
of particles is $\Delta N/N\sim \sqrt{N}/N \sim 10^{-10}$ which is
not important. However, for the condensate of fermionic atoms
$N\sim 10^5, ~~ \Delta N/N\ga 10^{-3}$ and this might be
physically significant.  An interesting open question is the role
of phase/particle number  uncertainty relation  $(\Delta
\theta)(\Delta N) \ga 1/2$ \cite{24} for condensate with  rather
low value of $N$. Another open question is the phase variation in
BEC-BCS crossover.

Now we return to the evaluation of the partition function. We
substitute (\ref{4})  into (\ref{3}) and rewrite the result in the
matrix      Nambu-Gorkov  representation. \be Z= \int D \psi^+
D\psi Dd^* D d  \exp \left\{ -\int^\beta_0 d\tau \int d\ver\left[
\frac{|d|^2}{g} +(\psi^+_{\sigma}
,\psi_{-\sigma})\left(\begin{array}{ll}
R&d\\d^*&R^T\end{array}\right) \left(\begin{array}{l}
\psi_\sigma\\\psi^+_{-\sigma}\end{array}\right) \right]\right\}
.\label{7}\ee Here $R=\left(\frac{\partial}{\partial
\tau}-\frac{\nabla^2}{2m}
-\mu\right),~~R^T=\left(\frac{\partial}{\partial
\tau}+\frac{\nabla^2}{2m} +\mu\right).$ Equation (\ref{7}) is
simply a mathematical construct completely equivalent to
(\ref{3})-(\ref{4}). In order to follow back the route from
(\ref{7}) to (\ref{3})-(\ref{4}) one should keep in mind the
anticommutativity of fermion fields and be ready to integrate by
parts the operator $R^T$. Next we integrate over the fermion
fields using the fact that (\ref{7}) contains a quadratic  form in
the exponent and the integral is of a Gaussian type. The result
reads \be Z=\int D d^* D d  \exp \left\{ -\frac{1}{g}\int^\beta_0
d\tau \int d\ver{|d|^2} -\beta \Omega_B (T,\mu, d,
d^*)\right\},\label{8}\ee where $\Omega_B$ is the Bogolubov
functional \be \beta \Omega_B =- tr ln \left( \begin{array}{ll}
R&d\\d^*&R^T\end{array}\right) = -\int^\beta_0 d\tau \int d\ver
\int \frac{d \vek}{(2\pi)^3} \frac{1}{\beta } \sum_{\omega_n} ln
(\omega^2_n + |d|^2 + E^2(k) ).\label{9}\ee

Here summation runs over the fermionic Matsubara modes $\omega_n
=\frac{\pi}{\beta} (2n+1)$ \cite{25}, and $E(k) =
\frac{k^2}{2m} -\mu$.

The value of the gap $d$ is given by the extremum condition for
the complete action in the exponent of (\ref{8}) \be
d=-\frac{g}{V_{3}} \frac{\delta  \Omega_B}{\delta
d^*},\label{10}\ee where $V_{3} =\int d\ver$. The Matsubara
summation in $\delta\Omega_B/\delta d^*$ is performed using the
relation \be T\sum_{\omega_n} \frac{2\varepsilon (k)}{\omega^2_n
+\varepsilon^2(k)} =th \frac{\varepsilon(k)}{2T}.\label{11}\ee
Here $T=1/\beta, \varepsilon^2 (k) =E^2 (k) + |d|^2$. From
(\ref{9}-\ref{11}) we obtain the self-consistency relation \be
1=\frac{g}{2} \int \frac{d\vek}{(2\pi)^3} \varepsilon^{-1}(k) th
\frac{\varepsilon(k)}{2T}.\label{12}\ee

At this point the BCS condition (iii) formulated above comes into
play. Attractive force between electrons due to phonons acts
within a   thin layer around the Fermi surface $-\omega_D \leq
E \leq\omega_D$. Therefore we may cut the integral in
(\ref{12}) and write \be \int \frac{d\vek}{(2\pi)^3}
\simeq\frac{mk_F}{2\pi^2}\int^{\omega_D}_{-\omega_D} d E
=2 N (0) \int^{\omega_D}_0 d E , \label{13}\ee where
$N(0)= mk_F/2\pi^2$ is the density of states on the Fermi surface.
Then we substitute (\ref{13}) into (\ref{12}), take the  limit $
T\to 0$ and  obtain the famous BCS solution \be d = 2 \omega_D
\exp \left(-\frac{1}{gN(0)}\right) = 2 \omega_D  \exp \left(
-\frac{\pi}{2k_F|a|}\right), \label{14}
 \ee where the scattering
length is introduced   according to $ g= 4\pi |a|/m$~~\footnote{
The rigorous definition of the scattering length  for the contact
interaction entering into (\ref{3}) may be found in
\cite{26,27}.}. The BCS regime corresponds to  $a<0$ if we use the
standard sign convention $k\cot \delta \to - a^{-1}$.  We can
estimate $|a|$ for a  typical superconductor with $\omega_D \sim
10^2 K$ and $d\sim T_c\sim $ few $K$. Then according to (\ref{14})
$gN(0)\simeq  0.3$ and $|a|\sim 10^{-8}$ cm $\sim k^{-1}_F.$

\section{Condensation in Quark Systems and in Atomic Fermi Gases}

With BEC and BCS ideas and technique at hand we may try to
understand to what extent are they applicable to the systems of
quarks and fermionic atoms.

Consider the system of $u$- and $ d$--quarks with density
(\ref{3}-\ref{5}) times larger than the density  of quarks in
normal nuclear matter (see Table 1). The corresponding value of
the chemical potential is around $\mu\simeq 0.4 $ GeV \cite{28}.
Such densities are typical for neutron stars. Strange quark starts
to  participate in pairing  at much  higher densities
corresponding  to $\mu\gg m_s \simeq 150$ MeV. Quark matter with
such moderate density got the name of 2 SC phase, where 2 stands
for the number of flavors and SC is  a shorthanded form of
"superconducting color". In enormous number of publications one
finds the statement that the BCS  regime really sets in starting
from $\mu\simeq 0.4$ GeV. This conclusion relies on the fact that
at such values of $\mu$ the gap equation of the type (\ref{10})
acquires a nontrivial solution (see \cite{28} and references in
\cite{8}). From the preceding discussion it should be  clear,
however, that a nonzero value of the gap is only a signal of the
presence of fermion pairs. Depending on the dynamics of the
system, on the fermion density, and on the temperature, such pairs
may be either stable, or fluctuating in time, may form  a BCS
condensate or dilute Bose gas, or undergo a Bose condensation. We
have seen that the BCS regime requires for its realization rather
special conditions (i)-(ii)-(iii) formulated above. Let us show
that these requirements are not met in the 2SC phase \cite{6,7}.

Scalar diquark in the 2SC regime is a would-be Cooper pair. We
recall that the spin structure of the order parameter in normal
superconductor has the form $d\sim \varphi_\sigma
(1)(i\sigma_2)_{\sigma\sigma'} \varphi_{\sigma'}(2)$. Similarly for
scalar diquark in $\bar 3$ color state we may write \cite{8}
$d_\alpha \sim \delta_{\alpha 3} \bar q^c_{\beta i}
\varepsilon_{\beta\gamma 3} (\tau_2)_{ij} \gamma_5 q_{\gamma j}$,
where the Greek indices stand for color, Latin ones for flavor,
$q^c = Cq^T,C$ is the charge conjugation operator, and the
presence of $\gamma_5$ makes the diquark a scalar (like $\sigma_2$
for electron pair). Diquark has a certain orientation in color
space (e.g., along the  third axis $\delta_{\alpha 3}$), therefore
$SU(3)_c$ gauge symmetry is broken, five of eight gluons are
massive. On the contrary, chiral symmetry is restored. The
symmetry pattern changes in  the ultra-high density region when
the strange quark enters the game. In the $\mu\to \infty$ limit
the real BCS regime sets in (though with some peculiarities
\cite{8}). We left out this topic since our main interest is
focused on the BCS-BEC interplay at moderate density.

 The dynamics of quark matter at moderate density
 is characterized by the  typical momenta $k\sim \mu \la 1$ GeV and by
 typical distances $r\sim \mu^{-1}\sim 1$ fm. In this domain
 theory faces the well known difficulties of nonperturbative  QCD
 and use is made of models like NJL or instanton gas. The form of
 the Bogolubov functional is practically independent of the model
 and reads \cite{29}
 \beeq \beta \Omega_B&=& -\frac12 tr ln \left( \begin{array}{cc}
 i\partial_\mu\gamma_\mu-i\mu\gamma_4&d\hat O\\ d^*\hat O^+&
 -i\partial^T_\mu\gamma^T_\mu + i\mu\gamma_4\end{array}\right)=\nonumber\\
&=& -tr ln (-\hat k-i\mu\gamma_4) -\frac12 tr ln \left\{ 1+
\frac{\hat O\hat O}{\hat k_+\hat k_-} |d|^2\right\},\label{15}\eeq
where $\hat k=k_\mu\gamma_\mu,~~ \hat O =\varepsilon_{\beta\gamma
3} (\tau_2)_{ij} C\gamma_5,~~ k_\pm = (\vek, k_4\pm i\mu)$.

The self-consistency relation (see(\ref{12})) at $T=0$ has the
form (for zero current quark masses) \be 1=\frac{g}{2}\int
\frac{d\vek}{(2\pi)^3} \{ \varepsilon^{-1}_+(k)
+\varepsilon^{-1}_- (k)\},\label{16}\ee where $\varepsilon^2_\pm
(k) = (k\mp\mu)^2+|d|^2.$

Acting essentially within the same scheme which led from (\ref{9})
to (\ref{14}) one obtains the result $d\simeq 0.1$ GeV almost
independently of the model \cite{1,2,8,28}. The roles of $N(0),
\omega_D$ and $E_F$ are played by $N(0)= 2\mu^2/\pi^2$, momentum
cutoff $\Lambda\simeq 0.7$ GeV, and $E_F \simeq \mu \simeq 0.5$
GeV. Let us compare these numbers with what we have in the
standard BCS picture \be BCS ~~d: \omega_D: E_F \simeq
1:10^2:10^4\label{17}\ee

\be 2SC ~~d: \Lambda: \mu\simeq :1:8:5.\label{18}\ee

We conclude that the hierarchy of scales inherent in the BCS
regime is badly broken in the 2SC phase. As we  have seen there is
another important parameter in the BCS scenario, namely
(see(\ref{2})) $k_F\xi\ga 10^3$. Let us estimate it for the 2SC
regime. Table 1 shows that $n^{1/3}\simeq 0.2$ GeV in the 2SC
phase\footnote{This value is obtained according to
$n=-\partial\Omega_{B}/\partial\mu$, but it is close to the result
for free degenerate quarks $ n=N_cN_f\mu^3/3\pi^2$.}. As for the
value of $\xi$, we may only rely on some estimates since rigorous
calculations are hardly possible in the nonperturbative QCD
region. One should also keep in mind   distinction between the
correlation length and the pair size. The two quantities coincide
in the BCS regime \cite{30}, while in BEC region the pair size is
smaller than the coherence length \cite{31,32}. The naive estimate
of $\xi$ may be done as follows. The energy spread of the
correlated pair of quarks is $\delta E\sim d$, quarks are
relativistic, hence $\delta p\sim d$, and therefore $\xi \sim
1/d\sim 2$ fm. Using the Klein-Gordon equation for the quark pair
\cite{33} one can obtain a better estimate $\xi \simeq  (\sqrt{3}
d)^{-1} \la 1$ fm\footnote{Recall that in the BCS theory
$\xi\simeq v_F/\pi d$ \cite{23}.}. Therefore we have $(k_F\simeq
\mu)$ \be k_F\xi \simeq 2.\label{19}\ee Comparing with (\ref{2})
we see that this value is about three orders of magnitude less
than in the BCS regime. The result (\ref{19}) indicates that the
2SC phase is in halfway between the BEC and BCS regions. The role
of fluctuations in the crossover domain is very important
\cite{34}. The significance of fluctuations is directly seen from
the estimate of the Ginzburg--Levanyuk parameter \be Gi \sim
(k_F\xi)^{-4}\sim 1/16.\label{20}\ee

For the ordinary superconductor $Gi\sim10^{-12}-10^{-14}$. Two
kinds of fluctuations should be considered -- that of the order
parameter $d$, and that of the gluon field $A$ \cite{35}. The
dominant ones are the gluon field  fluctuations since their
correlation length $T_g$ is about 0.2 fm \cite{36}, i.e., about 5
times smaller than the correlation length $\xi$ for $d$.
Fluctuation diamagnetism \cite{34} gives rise to the substantial
shift of the critical temperature \cite{37,38,39}\be T'_c=T_c
(1-g^2\xi^2\lan A^2\ran),\label{21}\ee where $g$ is the strong
coupling constant, $A$  -- the gluon field, $\lan ... \ran$ is the
average   over the quark fields, $T'_c\simeq 0.75 T_c$.

It is clear that further work beyond the  Mean Field Approximation
is needed  before the precise  dynamics of the 2SC phase is
finally elucidated.  In this  respect very helpful might be the
rapid progress taking place now in the study of similar problems
for ultracold fermionic atoms.  The central point here is the
experimental possibility to tune the interatomic scattering length
via the magnetic Feshbach resonance \cite{40,41}. The scattering
length entering into equation  (\ref{14}) may change its value and
sign as a function of the external magnetic field. Near the
resonance the dependence of the scattering length on the magnetic
field reads \be a(B)  =a_0
\left(1-\frac{\Delta}{B-B_r}\right),\label{22}\ee where  $a_0$ is
the ``background'' scattering length, $\Delta$ is some constant,
and $B_r$ is the value of the magnetic field corresponding to the
resonance at threshold $(a(B_r)\to \infty)$. The behavior  of  the
scattering length as a function of the  interaction strength $g$
is well known from quantum mechanics. With $g$ increasing the
scattering length first tends to $a\to -\infty$, then changes to
$a\to +\infty$ (i.e., pass through a resonance at threshold) and
then tends to  $a^{-1}\to \infty$ in the strong coupling limit
when the scattering length approximation is no longer legitimate.
In nuclear physics such an evolution may be regarded as a
transition from the $np$ virtual state with antiparallel spins and
$a\simeq -24$ fm to the deuteron with $a\simeq 5.4$ fm and
$\varepsilon_b \simeq 2.2$ MeV. In physics of cold fermion atoms
the region $a<0$ is called ``BCS side of resonance'' and the
region $a>0$ got the name ``BEC side of resonance''. We also know
that in the scattering length approximation the binding energy is
equal to $\varepsilon_b = (ma^2)^{-1}$ with $m$ being the mass of
each constituent forming the pair. At this point the ``connection
problem'' naturally arises: is it possible to connect by a certain
continuous transformation the BCS solution (\ref{14}) and the
scattering length solution presented above? Before we formulate a
positive  answer to this question a remark concerning
manipulations with the scattering length approximation  and with
the two-body interaction is general should be made. The BCS
solution (\ref{14}) is of essentially many-body nature. If we
consider an isolated pair  of electrons near the Fermi surface and
solve the eigenvalue problem in weak coupling limit and in thin
layer approximation, we would  obtain the result similar to
(\ref{14}) but with twice as large factor in the exponent. The
overlap of many Cooper pairs is important. The  solution of the
two-body problem should be embedded into the Mean  Field realm.
Therefore the scattering length $a$ should be understood as the
quantity renormalized by the medium. How to perform such  a
renormalization is a separate problem not discussed here.

The BEC   and BCS domains  are separated by the  ``unitary point''
$(k_F a)^{-1} =0$. In the vicinity of this point the scattering
length $a$ characterizes the size of the pair.   Our next task is
to demonstrate how the ``connection problem'' is solved, i.e.,
describe the BEC-BCS crossover.

\section{BCS-BEC Crossover}

The continuous evolution from BCS to BEC regime is called the
BCS-BEC crossover. Such a transition takes place either by
increasing  the  strength of the interaction or by  decreasing the
carrier density. The fact that the BCS wave function may undergo a
smooth evolution an describe the tightly   bound fermion pairs was
first noticed long ago \cite{42}-\cite{46}. According to
\cite{46}, the remark that ``there exists a transformation that
carries the BCS into BE state'' was originally made by F.J.Dyson
in 1957 (i.e., the same year that the BCS paper \cite{22} was
published). The BCS-BEC crossover for quarks was first discussed
in \cite{7}.

The explicit transformation from the BCS solution (\ref{14}) (weak
coupling) to the Schrodinger equation for  molecular-like state
(strong coupling) was performed in \cite{44}. Here we outline a
relativistic generalization of this procedure suited for light
quarks \cite{33}. We omit the details of the derivation since it
is similar to the derivation of the propagator described in
textbooks. One should take the derivative over $d^*$ of $ln Z_B$
(see (\ref{15})). Then at $T=0$ the positive and negative
frequencies relativistic wave functions in momentum space have the
form \be \varphi(k) =\frac{d(k)}{\varepsilon_+(k)},~~ \chi(k)
=\frac{d(k)}{\varepsilon_-(k)},\label{23}\ee where the quantities
$\varepsilon_\pm(k)$ are defined after Eq.(\ref{16}). The momentum
dependence of $d(k)$ means that we have assumed that the
four-fermion interaction (see (\ref{3})) is not point-like any
more. Then we use the self-consistency relation (\ref{16}) and
obtain the following set of coupled equations for $\varphi$ and
$\chi$ \be (\sqrt{k^2+m^2}-\mu) \varphi(k)
=(1-2n_k)\int\frac{d\vep}{(2\pi)^3} g(p-k) [\varphi(p)
+\chi(p)],\label{24}\ee

\be (\sqrt{k^2+m^2}+\mu)\chi(k) =(1-2\bar
n_k)\int\frac{d\vep}{(2\pi)^3} g(p-k) [\chi(p)
+\varphi(p)],\label{25}\ee \be 1-2n_k
=\frac{\sqrt{k^2+m^2}-\mu}{\varepsilon_+(k)},~~ 1-2\bar n_k
=\frac{\sqrt{k^2+m^2}+\mu}{\varepsilon_-(k)}.\label{26}\ee
 To obtain the Schrodinger equation we neglect the negative
 frequency component in (\ref{24}), expand the square root and
 consider the  dilute limit. The result is
 \be \left( \frac{k^2}{2m} -\tilde \mu\right )\varphi(k)=\int
 \frac{d\vep}{(2\pi)^3} g (p-k)  \varphi(p),\label{27}\ee
 where $\tilde\mu=\mu-m$ in line with the definition of the
 chemical potential in nonrelativistic case.

 In relativistic case the phase diagram in the $(n_k/\bar
 n_k,\mu$) plane has two symmetric   branches corresponding to
 quarks and antiquarks with possible ``exciton-like'' instability.

 In order to keep these two lectures down to a reasonable length we
  have not discussed the physics of ultra cold fermionic atoms at
  length. This subject is rapidly developing and one should follow
  the  current literature. The last remark we  wish to make
  concerns Eq. (\ref{22}) and Efimov effect \cite{47}. The point
  is that Feshbach  resonance may give rise to infinite number
  (proportional to $ln (|a|/r_0))$ of three body atomic states.

  These lectures were delivered at 8th Moscow School of Physics in
  early spring of 2005. The author wishes to express his deep
  gratitude to the Organizing Committee for the invitation and
 a  nice time spent there.

\end{document}